\documentclass[twocolumn]{aastex631}

\usepackage{amsmath,amssymb,bm}

\usepackage{comment}

\usepackage{cancel}
\usepackage{tikz}

\def \be{\begin{equation}}
\def \ee{\end{equation}}
\def \bea{\begin{eqnarray}}
\def \eea{\end{eqnarray}}

\hypersetup{
    urlcolor=cyan,
    }

\shorttitle{Little Red Dots as Quasi-stars}
\shortauthors{Begelman \& Dexter}

\graphicspath{{./}{figures/}}

\begin{document}

\title{Little Red Dots as Late-stage Quasi-stars}

\email{mitch@jila.colorado.edu}
\author[0000-0003-0936-8488]{Mitchell C. Begelman}
\affiliation{JILA, University of Colorado and National Institute of Standards and Technology, 440 UCB, Boulder, CO 80309-0440, USA}
\affiliation{Department of Astrophysical and Planetary Sciences, University of Colorado, 391 UCB, Boulder, CO 80309-0391, USA}

\author[0000-0003-3903-0373]{Jason Dexter}
\affiliation{JILA, University of Colorado and National Institute of Standards and Technology, 440 UCB, Boulder, CO 80309-0440, USA}
\affiliation{Department of Astrophysical and Planetary Sciences, University of Colorado, 391 UCB, Boulder, CO 80309-0391, USA}

\begin{abstract}
We argue that the ``Little Red Dots'' (LRDs) discovered with the {\it James Webb Space Telescope} are quasi-stars in their late stages of evolution.  Quasi-stars are hypothetical objects predicted to form following the core collapse of supermassive stars, and consist of black holes accreting from massive envelopes at a super-Eddington rate.  We show that models of late-stage quasi-stars, with black hole masses exceeding $\sim 10\%$ of the total, predict thermal and radiative properties that are insensitive to both black hole and envelope mass, and spectrally resemble LRDs. Specifically, we show that they are likely to exhibit reddish colors, a strong Balmer break, and possess conditions favorable to the production of Balmer lines that are broadened by electron scattering. Their huge electron column densities suppress any X-rays.  Late-stage quasi-stars, with black hole masses $\gtrsim 10^6 M_\odot$, should dominate the overall quasi-star population.  Their short predicted lifetimes (tens of Myr), coupled with the high observed comoving density of LRDs, suggest that most or all supermassive black holes go through a quasi-star/LRD phase during their formation and growth.
\end{abstract}


\section{Introduction}
\label{sect:intro}

Little Red Dots (LRDs) comprise a population of high-redshift objects discovered with the {\it James Webb Space Telescope} ({\it JWST}) that are characterized by a ``V-shaped'' spectrum in rest-frame UV-to-optical bands \citep{matthee24,greene24,kocevski24,taylor24,akins24}, with the transition occurring at the  rest-frame Balmer edge \citep{setton24,kokorev24,labbe24}.  The presence of broad ($\gtrsim 10^3$ km s$^{-1}$)  Balmer lines 
and a Balmer break stronger than can be readily accommodated with a stellar population \citep{inayoshi25,ji25,naidu25,degraaff25} suggest that the main source of emission is accretion onto a massive black hole (BH). Photoionization models \citep[e.g.][]{inayoshi24,naidu25} show that envelopes with density $n \sim 10^{11} \hspace{2pt} \rm cm^{-3}$, ionization parameter $U \sim 0.1$, and column density $N_H \sim 10^{26} \hspace{2pt} \rm cm^{-2}$ can produce the strong Balmer lines and break seen in LRDs.  The failure to detect X-rays from LRDs has been interpreted as evidence that high column densities associated with super-Eddington accretion  can suppress or obscure any coronal emission \citep{pacucci24,madau24,lambrides24,inayoshi24,rusakov25}.

These considerations have naturally led to speculation about the dynamical and energetic connections between the accreting black hole and envelope such as a ``black hole star'' \citep[BH$^*$:][]{naidu25} and analogous configurations supported by rapid infall from the environment \citep{rusakov25,kido25}.  Arguing inductively from observational and dynamical constraints, these proposals point out that LRD spectra can be reproduced with envelopes much less massive than the BH. In this paper, however, we take a deductive approach, placing LRDs in the context of a variant of the direct collapse model for supermassive black hole (SMBH) seed formation (e.g., \citealt{haehnelt93, umemura93, loeb94, eisenstein95a, madau01, bromm03, begelman06, lodato06, volonteri08, alexander14, shlosman16, natarajan17, pacucci17, wise19}).  We argue that LRDs could be ``quasi-stars'' seen in the latter stages of their evolution.  Quasi-stars are hypothetical objects that consist of BHs embedded in and accreting from gaseous envelopes with masses comparable to or greater than the BH mass. They were predicted to be the outcome of SMBH seed formation via core collapse of a supermassive star \citep[SMS:][]{begelman06}.  Since SMSs require accretion rates of $\gtrsim 0.1 M_\odot$ yr$^{-1}$ to form \citep{begelman10,hosokawa12,hosokawa13} and quasi-stars are predicted to last for tens of Myr, they would naturally acquire  masses of millions of solar masses toward the end of their lives.  For reasons we will discuss in \S \ref{sect:qsstructure}, we define a ``late-stage'' quasi-star to be one where the BH mass is $\gtrsim 0.1$ of the total mass (BH + envelope). 

Quasi-stars are powered by accretion onto the central BH at a rate that liberates the Eddington luminosity for the combined mass of the BH + envelope, which can greatly exceed $L_E$ for the BH alone.  As a result, the mass of a BH inside a quasi-star grows linearly with time instead of exponentially \citep{coughlin24}.  Therefore, a quasi-star spends most of its lifetime in the late stage of its evolution and we are most likely to see them in this stage.  But do they look like LRDs?  In \S \ref{sect:qsstructure} we review the recent theoretical work on late-stage quasi-stars by \cite{coughlin24}, contrasting it with earlier models that suggested the dispersal of the quasi-star envelope at lower BH masses.  We also extend this work to sketch the main characteristics of a quasi-star's radiative layer and photosphere.  In \S \ref{sect:qsatmosphere} we show that the resulting spectra may be consistent with observations of LRDs, including the red colors and Balmer break.  The broadening of the Balmer lines could be due to strong electron scattering between the thermalization layer and the photosphere, rather than Doppler broadening by bulk motion as a simple analogy with AGN broad-line regions would suggest.  We consider the evolutionary status of LRDs-as-quasi-stars in \S \ref{sect:evolution}, and discuss other potential observational signatures and summarize our argument in \S \ref{sect:summary}.

\section{Late-stage quasi-stars}
\label{sect:qsstructure}

\subsection{Internal structure}
\label{interior}

Early models of quasi-stars treated the case where the BH mass is a small fraction (typically $\lesssim 1 \%$) of the total mass, as may be the case shortly after core-collapse of the SMS \citep{begelman08}. The innermost region is treated as a modified Bondi accretion flow, through which the energy liberated by accretion percolates convectively.  Outside the Bondi radius, the quasi-star envelope is described as a convective ($\gamma = 4/3$) polytrope dominated by radiation pressure.  \cite{ball11} showed that these models fail when the BH mass $M_{\rm BH}$ exceeds $\sim 1 \%$ of the total quasi-star mass $M_*$, which is roughly the point at which the mass inside the Bondi radius becomes comparable to $M_{\rm BH}$. Citing the Sch\"onberg-Chandrasekhar limit, \cite{ball12} argue that the interior structure must become drastically different in order for quasi-stars to exist with higher mass ratios.

\cite{coughlin24} showed that the problems raised by \cite{ball11,ball12} can be addressed if the modified Bondi flow is replaced by an extended region of highly saturated convection, which carries a luminosity
\be
\label{Lsat}
L = 4\pi r^2 \beta p c_s, 
\ee
where $c_s = (4p/3\rho)^{1/2}$ is the sound speed associated with the radiation pressure $p$. The parameter $\beta$ represents the convective efficiency, such that $\beta \sim 1$ corresponds to energy flow at close to the sound speed. We expect that the net transport speed is less than this and adopt $\beta = 0.1$ as a fiducial value below, but note that the value is not theoretically well-constrained. Imposing hydrostatic equilibrium in this inner region implies 
\be
\label{eq:prho}
p \sim {L \over (GM_{\rm BH})^{1/2}} r^{-3/2} \qquad  \rho \sim {L \over (GM_{\rm BH})^{3/2}} r^{-1/2}.
\ee
Numerical models by \cite{coughlin24} show that the saturated convection zone transitions smoothly to weak convection at a radius ($r_i$) where the enclosed gas mass is comparable to the BH mass.  A zone of weak but efficient convection, modeled as a $\gamma = 4/3$ polytrope, can then be integrated outward until (formally) the density and pressure vanish or, more realistically, the convection becomes inefficient at a radius $r_r$\footnote{\cite{coughlin24} identified $r_r$ as the base of the ``radiative'' zone, but it is more accurately described as the base of an inefficient convection zone where the $\gamma = 4/3$ polytropic assumption fails. Because a small but  non-negligible fraction of the quasi-star mass (up to $\sim 10\%$) lies outside $r_r$, some form of convective energy transport must extend to larger radii.} where the radiative diffusion timescale across a pressure scale height $\Delta r$, $t_{\rm diff} \sim \rho \kappa (\Delta r)^2/c$, becomes shorter than the buoyancy timescale, $t_{\rm buoy} \sim (r_r^2 \Delta r/G M_r)^{1/2}$.  Here, $\kappa$ is the opacity, $M_r$ is the mass enclosed within $r_r$ and $\Delta r \sim (r_r^2/GM_r) (p/\rho)$. Making the reasonable assumptions that $M_r$ and $r_r$ are comparable to $M_*$ and $R_*$, the total quasi-star mass and radius, we find that the condition for inefficient convection reduces to equation (20) of \cite{coughlin24} (modulo the efficiency factor $\beta$) and we recover their quantitative results by assuming $\kappa \approx \kappa_{\rm sc}$, the electron scattering opacity.  We find, for example, that the BH should be able to grow to about 61\% of the total mass, at which point $r_i$ converges with $r_r$.

While the global properties of these models depend on $M_{\rm BH}$ and $M_*$ separately, they are strikingly insensitive to the ratio $M_{\rm BH}/M_*$ once  $M_{\rm BH} \gtrsim 0.1 M_*$  \citep[see][Fig.~6]{coughlin24}.  According to these models, the radius of a late-stage quasi-star is given by 
\be
\label{RQS}
R_* \approx 6 \times 10^{16} \beta_{-1}^{2/5} M_6^{3/5} \ {\rm cm}, 
\ee
where $\beta = 0.1 \beta_{-1}$ and $M_* = 10^6 M_6 M_\odot$.  Since the quasi-star luminosity must be close to the Eddington limit, $L_E = 1.4 \times 10^{44} M_6$ erg s$^{-1}$, the effective temperature is
\be
\label{Teff}
T_{\rm eff} \approx 3000 \beta_{-1}^{-1/5} M_6^{-1/20} \ {\rm K}. 
\ee
The ratio of gas pressure to radiation pressure is constant throughout the weakly convective region at a value $p_{\rm gas}/p_{\rm rad} \approx 3.3 \times 10^{-3} M_6^{-1/2}$, corresponding to an equation of state $p \approx 1 \times 10^{19} M_6^{2/3}\rho^{4/3}$ [cgs].\footnote{This equation of state corresponds to $p/\rho^{4/3} \equiv K \approx 1 G M_*^{2/3}$, which can be compared to the unique value for an $n=3$ polytrope, $K = 0.364 GM_*^{2/3}$.  The difference presumably arises from the different inner boundary conditions applied in the two cases.}

\subsection{Outer layers}
\label{exterior}

\cite{coughlin24} did not consider the structure of the quasi-star outside $r_r$, including the sub-photospheric layer where the spectrum is formed.  Throughout the efficient (polytropic) convective zone, the diffusive luminosity varies with radius as $L_{\rm rad}(r) = L_{E*}[M(r)/M_*](1 - p_g/p_r)$, where $L_{E*}$ is the Eddington limit for the entire mass $M_*$ (assuming constant opacity, which is reasonable for this argument), $M(r)$ is the enclosed mass at $r$, and $p_g/p_r$ is the fixed ratio of gas-to-radiation pressure.  Convection carries the difference between $L_{\rm rad}(r)$ and the total luminosity $L$, which is independent of $r$.  But what is the value of $L$?  Using our fiducial parameters, the fraction of quasi-star mass outside $r_r$ can be estimated to be $(\Delta M)_r/M_* \sim 0.1$. Therefore, if $L \approx L_{\rm rad}(r_r)$, then this is about 10\% lower than $L_{E*}$ because $\sim 10\%$ of the mass is located outside $r_r$.  So the radiation pressure would fall 10\% short of the value needed to support the outermost layers of stellar mass.  Gas pressure would have to take up the slack, but we have seen that $p_g/p_r \sim 0.003$ in the convective interior (for $M_6 = 1$), i.e., it is about 30 times too weak to provide the necessary support.
 
Therefore, $L$ cannot be as small as the local Eddington limit at $r_r$; it must be close to $L_{E*}$.  In this case, the radiative flux would exceed the Eddington limit near $r_r$. In perfect spherical symmetry, the outward radiation pressure gradient would have to be balanced by a steep {\it inward} gradient of gas pressure, creating a density inversion. In practice, this is probably very unstable, leading to vigorous mixing where gas is lifted into the outer layer by  the radiation force, becoming highly inhomogeneous \citep{dotan12}.  Leakage of radiation through low-density channels presumably relaxes the density inversion, and as a limiting case we might consider the mean density at $r_r \lesssim r \lesssim R_*$ to be roughly independent of radius.

Assuming uniform density, we can construct a toy model for the outer layers. The density is given by 
\be
\label{rhor}
\rho \sim \rho_r \approx 2.2 \times 10^{-13} \beta_{-1}^{-24/45} M_6^{-4/5} \ {\rm g \ cm}^{-3}, 
\ee
while the temperature and scattering optical depth at $r_r$ are 
\be
\label{Tr}
T_r \approx 1.5 \times 10^4 \beta_{-1}^{-8/45} M_6^{-1/10} \ {\rm K}, 
\ee
\be
\label{taur}
\tau_{{\rm sc},r} \approx \rho_r x \kappa_{\rm sc} \Delta r \approx 1.2 \times 10^3\beta_{-1}^{4/45} M_6^{-1/5} , 
\ee
where $\kappa_{\rm sc} = 0.34 x $ cm$^2$ g$^{-1}$ with $x \equiv n_e/n$ the ionization fraction.  Electron scattering overwhelmingly dominates the opacity throughout the layer, but there is sufficient absorption to maintain Local Thermal Equilibrium (LTE) near $r_r$.  

The condition for maintaining LTE is $\tau_{\rm abs} \tau_{\rm sc} \geq 1$ \citep{rybicki79} where $\tau_{\rm abs} = \rho \kappa_{\rm abs} \Delta r$ and $\kappa_{\rm abs}$ is the absorption opacity. Suppose there is some radius $r_{\rm LTE}$ inside the photosphere, where this condition is marginally satisfied.  The two main contributors to $\kappa_{\rm abs}$ are H$^-$ bound-free opacity and free-free absorption.  Because the ionization is so high at $r_{\rm LTE}$ (as we will see below), the H$^-$ abundance can be estimated  using the Saha equation for hydrogen alone and should not be sensitive to the presence or absence of trace metals.  We assume an H$^-$ photoionization cross-section of $3 \times 10^{-17}$ cm$^{-2}$, appropriate for the temperatures under consideration \citep{wishart79}. For free-free absorption we use a Kramers opacity $\kappa_{\rm abs} = 10^{23} \rho x^2 T^{-7/2}$ [cgs].  Using the Saha equation to obtain $x$ and noting that $T \propto p^{1/4} \propto \tau_{\rm sc}^{1/4}$ in the LTE zone, we can solve for the condition $\tau_{\rm abs} \tau_{\rm sc} = 1$, assuming (for simplicity) $\beta_{-1} = 1$.  For three representative quasi-star masses $M_6 = (0.1, 1, 10)$, we find $T_{\rm LTE} = (6100,  6000,  5950)$ K, $\tau_{\rm sc, LTE} = (5.8, 15,  39) $, and $x_{\rm LTE} = (0.21,  0.37,  0.64)$. H$^-$ absorption dominates at the LTE threshold for $M_6 = (0.1, 1)$; free-free absorption barely surpasses it for $M_6 = 10$.  Despite their similar temperatures, the surface layers of late-stage quasi-stars differ from the photospheres of lower main sequence stars and red giants in two fundamental ways. First, their low densities imply that electron scattering is the dominant opacity. This means that the outermost layers of late-stage quasi-stars depart from LTE and have color temperatures that are substantially higher than their effective temperatures --- unlike main sequence stars and red giants but similar to accretion disks in X-ray binaries \citep{shimura95,merloni00}. Second, the low density means that hydrogen can remain highly ionized down to temperatures as low as $\approx 6000$ K, whereas (for example) the much denser solar photosphere is extremely neutral.     

To summarize, the \cite{coughlin24} model of late-stage quasi-stars predicts a scattering-dominated photosphere with a color temperature of $\sim 6000$ K and an effective temperature $\sim 3000$ K, insensitive to both $M_*$ and $M_{\rm BH}$. The scattering optical depth of the non-LTE outer layer falls in the range $\tau_{\rm sc} \sim 5-40$ for $10^5 M_\odot < M_* < 10^7 M_\odot$. As in luminous accretion disks, the color correction factor is given by $\sim \tau_{\rm sc}^{1/4}$ in the limit $\tau_{\rm sc}^{1/4} \gg 1$.

\section{Quasi-star spectra}
\label{sect:qsatmosphere}

The combination of relatively low temperatures with high scattering optical depths of late-stage quasi-star atmospheres provides conditions conducive to reproducing several observed features of LRDs:

\begin{itemize}

    \item \underbar{\it Red colors}. Late-stage quasi-stars should have spectra similar to $\approx 6000 $ K blackbodies.  Using stacked data from \cite{akins24}, \cite{kido25} has shown that such spectra (with modest extinction) provide a reasonable fit, although a slightly cooler blackbody would work even better.  %

    \item \underbar{\it Strong Balmer break}. The Boltzmann equation for hydrogen in the marginal LTE zone gives a ratio of $n=2$ to ground state population of $n_2 / n_1 = 4 \exp (- 1.18 \times 10^5 /T) \approx 10^{-8}$ for $T = 6000$ K. Assuming a neutral fraction $x$, number density $10^{10} n_{10}$ cm$^{-3}$ and layer thickness $10^{16} \Delta r_{16}$ cm, the $n=2$ column density in the marginal LTE zone is $N_2 \approx 10^{18} (1-x) n_{10} \Delta r_{16}$ cm$^{-3}$.  Since the absorption cross-section at the Balmer edge is $\sim 10^{-17}$ cm$^2$, this is ample to produce a strong Balmer break if the gas is $\lesssim 90 \%$ ionized, or if the column density is higher --- both consistent with our models.
    
    \item \underbar{\it Balmer lines}.  The likelihood of forming a strong Balmer break goes hand-in-hand with the formation of Balmer lines in the non-LTE scattering layer. The large $n=2$ population means that Balmer lines are likely to remain optically thick well above the layer where continuum radiation goes out of equilibrium.  As the intensity of the continuum is diluted by scattering, the line intensity can be maintained at closer to its Planck value for the temperature, and thus stand out in emission. A more extreme limit of this process is thought to produce Balmer (and other) lines in the outer regions of cataclysmic variable accretion disks, which can be optically thin to continuum while being optically thick in the lines \citep{williams80,williams88}.   

    A number of LRD spectra show large Balmer decrements in the broad lines, H$\alpha$/H$\beta\gtrsim 9$, compared to the Case B recombination value $\sim 3$ \citep[e.g.][]{brooks25,torralba25,deugenio25}.  Such large ratios are typically explained by dust extinction, yet there are severe observational limits on $A_V$ \citep{casey24,setton25,chen25}.  The conditions in quasi-star atmospheres may be conducive to other mechanisms that can explain non-standard decrements, including collisional excitation and resonant line-scattering \citep{chang25,torralba25,deugenio25}.  The radial dependence of excitation levels in the non-LTE zone may also modify the Balmer decrements and help to explain the presence of both line emission and absorption \citep{matthee24}.

    \item \underbar{\it Scatter-broadened lines}. The Keplerian speed at $R_*$ is given by $v_* \approx 500 \beta_{-1}^{1/5} M_6^{1/5}$ km s$^{-1}$, too small to account for observed linewidths of $2000-3000$ km s$^{-1}$.  Electron thermal speeds in the scattering layer are $v_e \sim 400$ km s$^{-1}$, opening up the possibility that the linewidths are established by repeated scattering. Since the broadened linewidth is $\sim \tau_{\rm sc} v_e/c$, the optical depths estimated in our models would over-broaden the lines, and perhaps smear them out entirely if the line emission passed through the entire scattering layer.  \cite{rusakov25} found good fits to the exponential profile predicted for scatter-broadening in a sample of observed LRDs if $\tau_{\rm sc} \sim 2 $, several times smaller than our estimates of $\tau_{\rm sc, LTE}$.  We suggest that most of the line emission arises just below the scattering photosphere, where the mismatch between color temperature and brightness temperature in the continuum is greatest.  We note that the Balmer breaks observed in LRDs tend to be smeared over an extremely broad range of wavelengths, which could indicate that they are formed closer to the LTE layer and thus undergo a larger amount of scatter-broadening than the lines.

    \item \underbar{\it No X-rays}.  The large electron column densities in the scattering layer would eliminate any X-rays produced behind it.

\end{itemize}

Our intent in this section is to sketch (mainly) qualitatively a few predicted aspects of the late-stage quasi-star spectra that can be compared to LRD observations.  These need to be fleshed out with more detailed modeling.

\begin{figure*}
\begin{center}
    \includegraphics[width=0.95\textwidth]
    {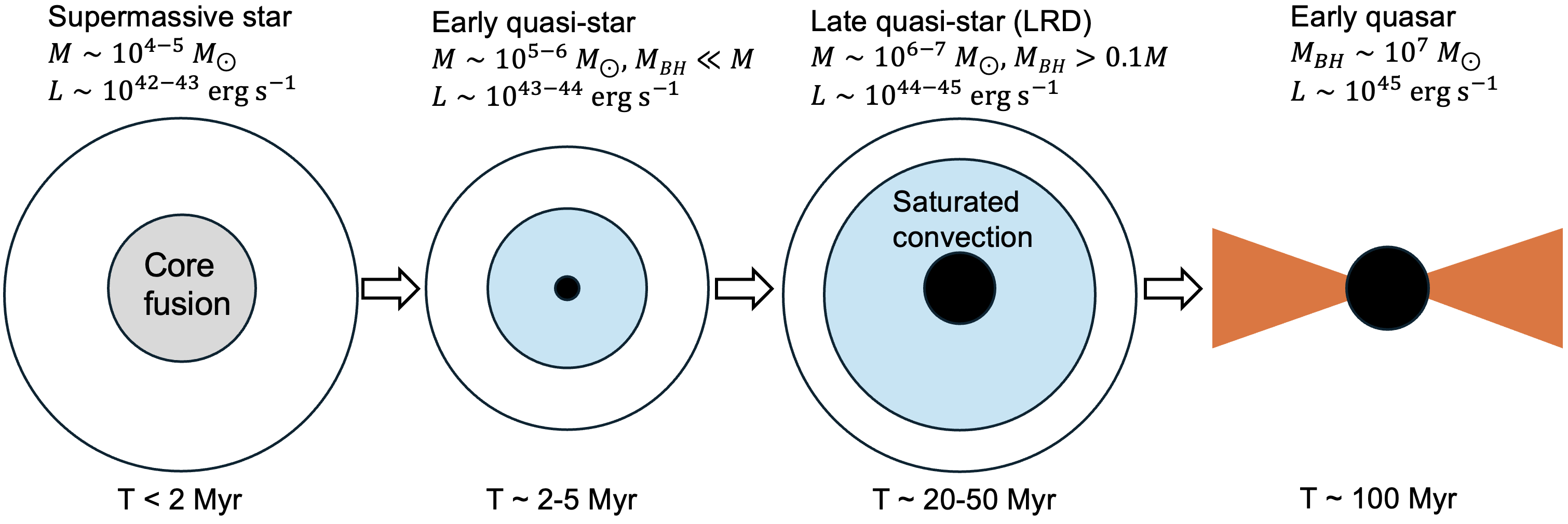}
    \end{center}
     \caption{Schematic diagram showing proposed stages in the formation and early growth of supermassive black holes from left to right. Approximate total ($M$) and black hole ($M_{BH}$) masses, luminosities $L$, and lifetimes $T$ are very rough estimates assuming a continued infall rate of $\gtrsim 0.1 M_\odot$ yr$^{-1}$. We propose that observed little red dots correspond to late quasi-stars. Rare precursors should appear bluer in color at similar or lower luminosity.}
     \label{fig:schematic_evolution}
\end{figure*}
  
\section{Evolutionary Context of Quasi-stars/LRDs}
\label{sect:evolution}
 
The time required for a BH to reach a mass fraction $\epsilon = M_{\rm BH}/ M_*$ inside a quasi-star is
\be
\label{massfrac}
t(\epsilon) \approx  \epsilon \eta_{-1} (39 - 77) \ {\rm Myr},
\ee
where $\eta = 0.1 \eta_{-1}$ is the BH accretion efficiency and  the lower value of the range corresponds to fixed $M_*$ while the higher value applies whenever the quasi-star envelope gains significant mass by accretion during this period \citep{coughlin24}.  \cite{coughlin24} argued that $\epsilon$ could not exceed 0.62 but  we cannot exclude the possibility that $\epsilon$ could be larger and might approach 1.  The linear growth rate implies that the number of quasi-stars seen at different evolutionary stages is roughly proportional to $\epsilon$, with late-stage quasi-stars being most common.

According to the proposed formation scenario \citep{begelman06,begelman10}, quasi-stars are set up following core collapse of an SMS (Fig.~\ref{fig:schematic_evolution}).  The ``prompt'' black hole mass following core collapse might only be a tiny fraction of the total SMS mass, depending on the internal angular momentum distribution of the star.  The energy liberated during the initial collapse is insufficient to unbind the SMS envelope, which instead expands and settles into a convective configuration \citep{begelman08} with the internal accretion rate onto the BH adjusting so that it liberates the Eddington limit corresponding to the mass of the entire envelope, not just that of the central BH.   

A mass supply rate of $\dot M \gtrsim 0.1  M_\odot$ yr$^{-1}$ from the local environment is a necessary condition to create an SMS, which increases in mass on a timescale shorter than its Kelvin-Helmholtz time until hydrogen ignites in its core \citep{begelman06,hosokawa12,hosokawa13}.  Although ignition occurs at a relatively modest stellar mass $\sim 10^3 M_\odot$ \citep{begelman06,begelman10}, by the time core collapse occurs after $\sim 2$ Myr the total mass has increased to $\sim 2 \times 10^5 \dot M_{-1} M_\odot $, where $\dot M = 0.1 \dot M_{-1} \ M_\odot $ yr$^{-1}$.  If the initial mass supply rate is steady for tens of Myr then most quasi-stars and their central BHs should reach masses of at least several million $M_\odot$ by their late stages.  The final masses could be lower if the mass supply is intermittent on shorter timescales or if there is a high rate of mass loss \citep{fiacconi16}.  On the other hand, there could be systems with inflow rates substantially larger than the minimum required, leading to quasi-star and BH masses $\gtrsim 10^7 M_\odot$.  For our discussion below, we will assume that the $\dot M$ distribution is skewed to smaller values and that the typical late-stage quasi-star has a mass of a few times $10^6 M_\odot$ up to $\sim 10^7 M_\odot$ and a bolometric luminosity of $\sim 10^{44}-10^{45}$ erg s$^{-1}$. 

Although LRDs are primarily found at high redshifts, this process might be triggered any time the requisite mass flux pours into a protogalactic nucleus that does not already contain a massive BH.  Thus, the recent discovery of several LRDs at $z =0.1 - 0.2$ \citep{lin25,ji25b} could indicate the existence of rare, compact dark matter halos that have only received their first rapid infall of baryons relatively recently.

At their typical redshift of $z \sim 5-6$, LRDs have a comoving density of $\sim 10^{-5}$ cMpc$^{-3}$, making them $\sim 100$ times less common on the sky than UV-selected galaxies \citep{matthee24,greene24}.  However, if they are quasi-stars then their typical lifetime is $\lesssim 50$ Myr, i.e.~$\lesssim 5\%$ of the age of the universe at $z \sim 5$.  Their observed density would then imply that LRDs must be remarkably common as a physical phenomenon: at some point, a significant fraction of all high-redshift galaxies must host an LRD.  This suggests that LRDs --- and by inference, SMSs and quasi-stars --- are ubiquitous features of SMBH formation and possibly the main channel.

The BH mass density left behind by LRDs/quasi-stars at $z \sim 5$ is then $\sim 250 M_6 M_\odot$ cMpc$^{-3}$.  If a typical LRD black hole subsequently grows (as an AGN) by a factor of $\sim 100 M_6^{-1}$ following its quasi-star epoch to become an SMBH with a median mass $\sim 10^8 M_\odot$, then the BH mass density resulting from LRD/quasi-star seeds is $\sim 2.5 \times 10^4 M_6 M_\odot$ cMpc$^{-3}$, only a few times smaller than the present-day BH mass density \citep{yu02}.  Several factors, including shorter lifetimes, could increase the contribution of LRDs to the present-day BH population.  A ubiquitous quasi-star/LRD stage, in which SMBHs acquire their first $\sim 10^6 - 10^7 M_\odot$ of mass in only a few 10's of Myr, could have important implications for galaxy formation.  Because this growth is first smothered within the quasi-star, feedback is relatively ineffective and only affects star formation as the BH grows further, as an AGN, within a more massive merged host galaxy.   
 
\section{Discussion}
\label{sect:summary}

Our proposed identification of LRDs with late-stage quasi-stars is based on two arguments: 1) the similarity of predicted quasi-star atmospheres with the radiating layers inferred to exist in LRDs from their observational properties \citep[e.g.][]{inayoshi25, ji25,naidu25,degraaff25,kido25}; and 2) the fact that quasi-stars are a natural prediction of SMBH seed formation through the core collapse of SMSs.  LRD radiative models already in the literature posit relatively low-mass envelopes associated with an accretion flow or supported in some other way (or assumed by fiat), but the mass estimates obtained are similar to the masses of the {\it observable} part of our massive quasistar envelopes.  For example, the non-LTE layer in our model, which sets the color temperature and Balmer features, has a mass of only $\Delta M_{\rm NLTE} \approx (63, 2700, 6.3 \times 10^4) M_\odot$ for $M_* = (10^5, 10^6, 10^7)M_\odot$, corresponding to fractional mass $\Delta M_{\rm NLTE} / M_* \sim (0.63, 2.7, 6.3)\times 10^{-3}$.  But the extant models differ from the quasi-star model conceptually in that the radiating layer in the quasi-star model forms a thin veneer overlying an extremely massive envelope that determines the radiating layer's properties.  Rather than thinking of this layer as a kind of AGN broad line region photoionized by a central source, we argue that it is more appropriate to think of it as a convectively powered envelope producing its own radiation in situ \citep[as in the model by][]{kido25} --- like a true star.  

In detail, we find other important differences between the quasi-star picture and other models.  The quasi-star radius is significantly larger than radii assumed in other models, by up to an order of magnitude.  The radial scale is set by convective processes in the deep interior and cannot be chosen arbitrarily, although refinements of the convective modeling --- which may change the radius somewhat --- are certainly warranted.  Another key feature we have emphasized --- partly the result of the large radius --- is that our radiating layers are highly ionized and dominated by electron scattering, despite their relatively low temperatures.  This means that the color temperatures of quasi-star/LRDs can be a factor $\gtrsim 2$ times larger than their effective temperatures.
 
From an evolutionary perspective, the most important implication of our model is that LRDs should be short-lived --- the central BH can swallow the entire envelope (or as much of it as is allowed) in roughly one Salpeter time, modulo an uncertain numerical factor.  This is true regardless of the initial BH mass, because the accretion rate is tied to the Eddington limit for the entire envelope.  Unless the external mass supply rate $\dot M$ exponentiates with time to keep pace, which seems unlikely, the Eddington limit of the BH will outrun the mass supply and the LRD/quasi-star phase will end.  Thus, the small observed number of LRDs compared to UV-selected galaxies does not mean that LRDs are physically rare, but quite the opposite: essentially every galaxy must host an LRD at some point.  We therefore propose that LRDs are an integral stage of the formation and growth process of supermassive black holes.

As we have noted, low-mass gaseous envelopes that reproduce the observed properties of LRDs could be difficult to distinguish from the atmospheres of quasi-stars --- some mechanisms for producing such envelopes  have been discussed in the literature \citep[e.g.][]{begelman12b,kido25,liu25}.  Such stand-alone envelopes could thus provide an alternative to the model discussed here.  However, the large inferred population of quasi-star/LRDs puts pressure on how important these processes could be for producing additional LRD episodes after the end of the quasi-star stage. 

The picture of LRDs as quasi-stars carries with it an evolutionary sequence that includes stages prior to and following the phenomena we have focused on in this paper.  We have illustrated this sequence schematically in Fig.~\ref{fig:schematic_evolution}.  The SMS stage should last only $\sim 2-5 \%$ of a late-stage quasi-star lifetime and should thus be rarest on the sky.  They should also be $\sim 10-100$ times less luminous on average, i.e. $L \sim 10^{42}-10^{43}$ erg s$^{-1}$, with color temperatures ranging from $\sim 5000$ K at lower masses to $\sim 10^4$ K as masses approach $10^5 M_\odot$ \citep{hosokawa13}.  Early-stage quasi-stars should be more common than SMSs ($\sim 10 \%$ compared to late-stage quasi-stars) and --- at constant envelope mass --- should evolve from hotter (several $\times 10^4$ K) to cooler color temperatures as the BH grows inside \citep{coughlin24}. However, this evolution could be obscured by the diversity of quasi-star masses and mass ratios. Efforts should be made to find the immediate precursors to LRDs.  Little is known about the fate of a quasi-star once the BH swallows most or all of the envelope, e.g., whether the residual gas disperses or is able to settle into a steady state with a much lower envelope mass if matter continues to be supplied from the environment at a sufficient rate \citep[e.g., resembling one of the stand-alone models as in][]{naidu25,degraaff25,kido25}.  Alternatively, if stripped of their envelopes these immediate post-quasi-stars could be the first proper AGNs in the universe --- examples of which may have already been discovered \citep[e.g.][]{bogdan24,natarajan24}.

\begin{acknowledgments}
We acknowledge support from NASA Astrophysics Theory Program grants 80NSSC22K0826 and 80NSSC24K1094.  We thank Eric Coughlin and Erica Nelson for helpful discussions, and the anonymous referee whose recommendations helped us to improve the paper.
\end{acknowledgments}

\bibliography{biblio}{}
\bibliographystyle{aasjournal}
 


\end{document}